\documentclass{article}
\usepackage{amsmath}
\usepackage{amssymb}
\usepackage{graphicx}
\usepackage{epsfig}
\usepackage{bm}

\def\spose#1{\hbox to 0pt{#1\hss}}
\def\simlt{\mathrel{\spose{\lower 3pt\hbox{$\mathchar"218$}}
     \raise 2.0pt\hbox{$\mathchar"13C$}}}
\def\simgt{\mathrel{\spose{\lower 3pt\hbox{$\mathchar"218$}}
     \raise 2.0pt\hbox{$\mathchar"13E$}}}
\def\id{\mathbb{I}}

\def\p{\mbox{\boldmath$\displaystyle\mathbf{p}$}}

\def\bv{\mbox{\boldmath$\displaystyle\mathbf{\varphi}$}}
\def\bt{\mbox{\boldmath$\displaystyle\mathbf{\theta}$}}

\def\0{\mbox{\boldmath$\displaystyle\mathbf{0}$}}
\def\s{\mbox{\boldmath$\displaystyle\mathbf{\sigma}$}}

\begin{document}

\textbf{\large Towards a relativity of dark-matter rods and clocks}

{\sc D. V. Ahluwalia}\\ 
\textit{\small Department of Physics and Astronomy\\
Rutherford Building, University of Canterbury \\ 
Private Bag 4800, Christchurch 8020, New Zealand}
\\ \\
\textit{E-mail: dharamvir.ahluwalia@canterbury.ac.nz}

 \begin{abstract}  
 In the absence of dark matter, the dynamical and kinematical
 interpretations of the special relativistic spacetime have been and
 still are the topic of philosophic debate, which whilst fertile, is
 by and large of little predictive power. This changes dramatically 
 if the debate includes a dark matter candidate in a ``non-trivial''
 extension of the standard model. Here I argue that rods and clocks
 made out of dark matter may not reveal the same underlying algebraic
 structure as the rods and clocks made out of standard model
 particles. For the sake of concreteness I here exemplify the argument
 by looking at a particular dark matter candidate called
 Elko. Inevitably, one is led to the conclusion that gravity within
 the dark sector, and at the interface between dark matter and
 standard-model matter, may deviate from the canonical general
 relativistic predictions. For Elko dark matter such effects will be
 of second order in the sense that they will depend only on the
 angular momentum and spin of the gravitational environment.
 \end{abstract}
 
\vspace{52mm}
\begin{quote}
{\sc Essay written for the Gravity Research Foundation 2009 Awards for
Essays on Gravitation; submission date: 31 March 2009; Selected
for an ``honorable mention'' 15 May 2009.}
\end{quote}
\newpage
 In the dynamical interpretation of the spacetime of special
 relativity (SR), Lorentz algebra ($\mathcal{L}$) emerges as a
 property of the standard-model (SM) fields, and not as an intrinsic
 attribute of spacetime itself~\cite{Brown:2005hr}. The Lorentz
 algebra describes not only the spinorial characteristics of the SM
 matter, but also the four-vector character of the Einsteinian
 spacetime. That is, rods and clocks made of SM material reveal the
 symmetries underlying the material they are made of, rather than
 other way around where spacetime is a kinematic element and its
 symmetries dictate what matter fields can exist.

 However, as long as there was no inkling of dark matter, the
 dynamical interpretation of spacetime and its kinematical counterpart
 were entirely a matter of philosophic debate. However, as shall be
 argued here, recent work on the quantum field theory of Elko dark
 matter suggests a logical corollary to the dynamical
 proposition~\cite{Ahluwalia:2004sz,Ahluwalia:2004ab,Boehmer:2006qq,daRocha:2007pz,Gredat:2008qf,daRocha:2005ti,daRocha:2008we,daRocha:2009gb}:
\textit{rods and clocks made out of dark matter may not reveal the same underlying algebraic structure as do rods and clocks made out of the standard model particles.}

 The proof of this corollary is subtle and requires a self contained
 technical presentation of the Elko quantum field. For the sake of
 expediency we shall here confine to a heuristic argument that may
 suffice to convince the reader that the present suggestion is a valid
 one.

\textbf{Outline of an algebraic derivation of special relativistic
boosts and rotations.}

 First let us take a brief walk through the familiar territory of the
 SM matter fields and the realm of SR. In the process we shall arrive
 at the outline of an algebraic derivation of special relativistic
 boosts and rotations. The task of this excursion is to show with an
 element of transparency as to how the symmetries of the underlying
 matter determines the properties of the emergent spacetime. This
 gives the dynamical interpretation of special relativity a concrete
 form and allows us to suggest the said corollary.

 All standard model matter fields are described by the Dirac quantum
 field. Since our immediate interest is to construct the classical,
 rather than a quantum spacetime, we will only consider the Dirac
 spinors.\footnote{The presented exercise is also suggestive of a new
 procedure for constructing a quantum spacetime.} These four-component
 spinors are a direct sum of the massive two-component Weyl spinors of
 the right and left-handed type. In an argument spanning several
 chapters, Weinberg shows that the Dirac rest spinors must have a very
 specific form in which the left-transforming and the
 right-transforming Weyl components of the Dirac spinors carry
 \textit{matching} helicities~\cite{Weinberg:1995mt}. He further
 arrives at the conclusion that the global phase associated with these
 spinors \textit{cannot} be arbitrary, but must take a set of very
 specific values in order that the field operator transforms unitarily
 between the inertial frames of SR, and in order that locality be
 preserved.\footnote{This is not well known, and numerous textbooks
 suffer from errors that can be traced back to this neglect. Among the
 few books that do not suffer from this ailment the reader may consult
 Srednicki's classic~\cite{Srednicki:2007qs}.} Thus the special
 relativistic demand of locality is incorporated in the description of
 SM fermions through a very specific choice of phases in the SM matter
 fields. This is beyond the naive assumption that Dirac spinors chosen
 simply as solutions of the Dirac equation will yield a local quantum
 field that corresponds to a unitary representation of the inhomogeneous
 Lorentz group.

 To further understand the relation between Dirac spinors and SR
 spacetime, note that the boost transformation for the right-handed
 Weyl spinors is associated with the boost generator $-i \s/2$ and
 reads\footnote{Most of the notation is standard, and is defined in
 references~\cite{Ahluwalia:2004sz,Ahluwalia:2004ab}. In particular,
 $\bv$ is the boost parameter, and $\bt$ represents the rotational
 parameter. The $\s$ are the standard Pauli matrices.}
	\begin{equation} \kappa_r = \exp\left(+
	\frac{\s}{2}\cdot\bv\right) = \frac{1}{\sqrt{2 m (E +
	m)}}\left[(E+m) \id + \s\cdot\p\right] 
	\end{equation}
 Similarly, the boost transformation for the left-handed Weyl spinors
 is associated with the boost generator $+ i \s/2$ and is
	\begin{equation}
	\kappa_\ell = \exp\left(-\frac{\s}{2}\cdot\bv\right) = 
	\frac{1}{\sqrt{2 m (E + m)}}\left[(E+m) \id - \s\cdot\p\right]
	\end{equation}
 Analogous expressions can be written for the spinorial transformation
 matrix for rotations. These can be obtained by taking note that both
 the right and the left-handed Weyl spinors are associated with the
 rotational generator $\s/2$. The rotational transformation matrix for
 Dirac spinors is simply the \textit{direct sum}
 $\zeta_{Dirac}:= \zeta_r\oplus\zeta_\ell$, where
 	\begin{equation}
	\zeta_r = \zeta_\ell = \exp\left(i \frac{\s}{2}\cdot\bt\right)
	\end{equation}
 The boost operator for the Dirac spinors is the \textit{direct sum}
 $\kappa_r\oplus\kappa_\ell$.

 This is all very familiar. In fact, had
 one arrived at $\mathcal{L}$ through experiments on electrons, one
 could have arrived at the relativity of spacetime in a more
 transparent way; or at least in a manner that manifestly reflects the
 emergence of spacetime from the underlying matter fields. To
 illustrate, let us take the convention where the spacetime four
 vectors are represented as $x^\mu:=\{t,x,y,z\}$. Then the boost and
 rotation matrices of SR in terms of $\kappa_r, \kappa_\ell, \zeta_r$,
 and $\zeta_\ell$ read
	\begin{equation}
	\alpha \left[\kappa_r\otimes\kappa_\ell\right] \alpha^{-1},\qquad
	\alpha \left[\zeta_r\otimes\zeta_\ell\right] \alpha^{-1} 
        \label{eq:SR}
	\end{equation}
 with $\alpha$ given by
	\begin{equation}
	\alpha = \sqrt{\frac{1}{2}}\left(\begin{array}{cccc}
	0 & i &-i &0\\
	-i & 0 & 0& i \\
	1 & 0 & 0 & 1\\
	0 & i & i & 0
	\end{array}\right)
	\end{equation}
 The reader may wish to verify that the above two expressions coincide
 with the Einsteinian results for SR. The generators associated with
 $\kappa_r\oplus\kappa_\ell$ and $\zeta_r\oplus\zeta_\ell$ satisfy the
 Lorentz algebra, as do those associated with
 $\kappa_r\otimes\kappa_\ell$ and $\zeta_r\otimes\zeta_\ell$. The need
 for $\alpha$ is simply to respect the standard co-ordinate convention.

 \textit{This brief review and outline thus serves to bring out the
 algebraic harmony between the spinorial properties of the SM matter
 fields and the Einsteinian spacetime of SR.}  The important
 observation is that for the Dirac spinors both the right- and left-
 Weyl components carry \textit{same} helicity.

 \textbf{Towards a relativity of Elko dark-matter rods and clocks.}

 The starting point of the Elko quantum field is a complete set of
 eigenspinors of the spin one half charge conjugation operator,
 $\mathcal{C}$.  These are obtained from the observation that if
 $\phi_\ell(\p)$ transforms as a \textit{left}-handed (massive) Weyl spinor
 then (a) under a Lorentz boost, $\eta \Theta \phi_\ell^\ast(\p)$
 transforms as a \textit{right}-handed Weyl spinor (where $\eta$ is an
 unspecified phase to be determined below and $\Theta$ is Wigner's
 time reversal operator for spin one half,
 $\Theta\left[\s/2\right]\Theta^{-1} = - \left[\s/2\right]^\ast$); and
 (b) the helicity of $\eta \Theta \phi_\ell^\ast(\p)$ is {\em opposite} to
 that of $\phi_\ell(\p)$. The phase $\eta$ is fixed to be $\pm i$ by
 demanding that
	\begin{equation}
	\chi(\p)= \left(\begin{array}{c}
 	\eta \Theta \phi_\ell^\ast(\p)\\
	\phi_\ell(\p)
	\end{array}\right)\label{eq:taup}
	\end{equation}
 be eigenspinors of $\mathcal{C}$ with eigenvalues $\pm 1$. A simple
 calculation shows that unlike the 
 Dirac spinors of the SM, the $\chi(\p)$ are not eigenspinors of the
 $\gamma_\mu p^\mu$
 operator~\cite{Dvoeglazov:1995eg,Ahluwalia:2004ab}; and hence do not
 satisfy Dirac equation. This, along with the results that followed,
 had a series of potentially important theoretical and
 phenomenological consequences. Our excitement is noted in the opening
 and concluding lines of the \textit{abstract} of our PRD paper. These
 read~\cite{Ahluwalia:2004sz}

 \begin{quote} We report an unexpected
 theoretical discovery of a spin one-half matter field with mass
 dimension one. ... Its dominant interaction with known forms of
 matter is via Higgs, and with gravity. This aspect leads us to
 contemplate it as a first-principle candidate for dark matter.
 \end{quote} 
 The details appeared in a 2005 JCAP paper~\cite{Ahluwalia:2004ab}
 with additional insights reported in~\cite{Ahluwalia:2008xi}.

 The \textit{uniqueness} of the Dirac quantum field -- modulo the 1937
 observation of Majorana~\cite{Majorana:1937vz} -- within the context
 of the inhomogeneous Lorentz spacetime symmetries is apparent from the
 noted work of Weinberg. This immediately raises the question as to
 how the Elko quantum field be accommodated within inhomogeneous
 Lorentz symmetries. The rough answer is that just as certain
 solutions of a wave equation need not exhibit the symmetries
 underling the wave equation, certain quantum fields may not carry the
 full symmetries of the underlying algebraic structure -- the
 violation may be similar to spontaneous symmetry breaking in the SM
 physics, or something not yet fully known. The Elko construct demands
 that the Lorentz spacetime symmetries must be violated, or modified,
 without changing the generators from their special
 relativistic form.

 \textbf{Concluding remarks.}

  The Elkonian special relativity necessarily suggests an as-yet-unknown
 modification of general relativity for the Elkonian dark sector
 and its interface with the SM matter and gauge fields. It is already
 apparent that these deviations from GR shall be second order. Second
 order, in the sense that these shall depend on the angular momentum
 and the spin content of the gravitational environment with as yet to
 be deciphered consequences for astrophysics and cosmology. Yet,
 these effects may already be present in the high-precision cosmic
 microwave background
 data~\cite{Samal:2007nw,Samal:2008nv,Hinshaw:2008kr}.

 On the more urgent side, understanding the violation of the 
 canonical rotational symmetry by the Elkonian dark sector 
 remains an outstanding preoccupation. Insights in this regard 
 are emerging from the work reported in this essay.

 The arguments that I have outlined here are far from
 complete. However, I hope this essay serves the purpose of convincing
 the reader that the rods and clocks made out of dark matter -- which
 need not be Elko -- may not reveal the same underlying algebraic
 structure as the rods and clocks made out of the standard model
 particles.

Since this essay was first written, we have discovered that 
 the said violation of the Lorentz symmetries happens  through the introduction of  a preferred axis along which Elko quantum field becomes local. This work shall soon be reported in a separate publication.

 \textbf{Acknowledgements.} It is my pleasure to thank Adam Gillard,
 Cheng-Yang Lee, Dmitri Schritt and Ben Martin for our continuing
 discussion on Elko dark matter. I also thank Karl-Henning Rehren for
 his comments on Elko, and an anonymous referee of JCAP.   Finally,
 it is my pleasure to acknowledge lengthy discussion with Harvey Brown
 on the dynamical interpretation of spacetime. These were
 important for the view taken in this essay.

\providecommand{\href}[2]{#2}\begingroup\raggedright\endgroup

 \end{document}